\documentclass[conference]{IEEEtran}
\IEEEoverridecommandlockouts
\usepackage{cite}
\usepackage{amsmath,amssymb,amsfonts}
\usepackage{graphicx,setspace}
\usepackage{textcomp}
\usepackage{xcolor}
\usepackage{algorithm,algpseudocode}
\usepackage{textcomp}
\usepackage{array}
\usepackage{tabularx}
\usepackage{booktabs}
\usepackage{mathtools}
\usepackage{comment}
\usepackage{amsthm}
\usepackage{url}

\usepackage{epstopdf}

\usepackage{subfigure}
\DeclarePairedDelimiter{\norm}{\lVert}{\rVert}

\theoremstyle{definition}

\newtheorem{remark}{Remark}

\allowdisplaybreaks
\begin{document}

\title{On the Optimal 3D Placement of a UAV Base 
Station for Maximal Coverage of UAV Users\\
\thanks{The authors would like to acknowledge the support of the Natural Sciences and Engineering Research Council Canada (NSERC), Huawei Canada, and Tunisian Ministry of Higher Education and Scientific Research.}}

\author{\IEEEauthorblockN{  Nesrine~Cherif\IEEEauthorrefmark{1}, Wael Jaafar \IEEEauthorrefmark{2}, Halim Yanikomeroglu\IEEEauthorrefmark{2}, and Abbas Yongacoglu\IEEEauthorrefmark{1} \\
	\IEEEauthorblockA{\IEEEauthorrefmark{1}School of Electrical Engineering and Computer Science, University of Ottawa, Ottawa, ON, Canada.\\
		\IEEEauthorrefmark{2} Department of Systems and Computer Engineering, Carleton University, Ottawa, ON, Canada.	\\
		Email:	\IEEEauthorrefmark{1}\!\{ncher082, yongac\}@uottawa.ca,	\IEEEauthorrefmark{2}\!\{waeljaafar, halim\}@sce.carleton.ca.}}
 \vspace{-2em}
}
\maketitle
\linespread{1.1}
\begin{abstract}
Unmanned aerial vehicles (UAVs) can be users that support new applications, or be communication access points that serve terrestrial and/or aerial users. In this paper, we focus on the connectivity problem of aerial users when they are exclusively served by aerial base stations (BS), i.e., UAV-BSs. Specifically, the 3D placement problem of a directional-antenna equipped UAV-BS, aiming to maximize the number of covered aerial users under a spectrum sharing policy with terrestrial networks, is investigated. Given a known spectrum sharing policy between the aerial and
terrestrial networks, we propose a 3D placement algorithm that
achieves optimality. Simulation results show the performance of
our approach, in terms of number of covered aerial users for
different configurations and parameters, such as the spectrum
sharing policy, antenna beamwidth, transmit power, and aerial
users density. These results represent novel guidelines for exclusive aerial networks deployment and applications, distinctively for orthogonal and non-orthogonal spectrum sharing policies with terrestrial networks.
\end{abstract}

\section{Introduction}
\IEEEPARstart{T}{he} number of {unmanned} aerial vehicles (UAVs) is expected to skyrocket in the near future according to reports from the US federal aviation administration (FAA). Based on statistics of March 2020, slightly over 1.5 million UAVs are registered in the US, and the number is projected to increase exponentially in the upcoming years, as UAVs are becoming an important part of the every-day's life \cite{Faawebsite}. UAVs are expected to be part of the 5G wireless networks in two ways: 1) as an aerial base station (UAV-BS) that provides cellular connectivity \cite{irem}  
or 2) as an aerial user (UAV-UE) \cite{mozaffari2018beyond}. 
Due to their flexible deployment, UAV-BSs (a.k.a., UxNBs in 3GPP terminology \cite{3gppTS22125}), can provide in-demand wireless broadband in temporary events or ubiquitous cellular coverage in hard-to-access remote areas  \cite{irem}. 
Meanwhile, UAV-UEs are used to accomplish new services and tasks, such as flying taxis, package delivery, video surveillance, aerial security inspection, wireless sensors data collection, etc. To do so, UAV-UEs require a reliable connectivity 
to the wireless network. 
Since most terrestrial BSs are down-tilted, their vertical coverage is rather patchy and does not satisfy the stringent requirements of UAV-UEs operations \cite{3gpp777,nesrineglobecom}. In addition, hard-to-access remote areas may not have terrestrial BSs, hence, UAV-UEs would be left without any cellular connectivity to fulfill their tasks. 
Consequently, providing ubiquitous aerial coverage using UAV-BSs becomes a key enabler of reliable UAV-UEs communications. Hence, a comprehensive aerial coverage planning using UAV-BSs is urgently needed to enable UAV-UEs based services.

Several works on UAV-based wireless communications have 
investigated challenges linked to the efficient deployment of UAV-BSs to serve terrestrial users. For instance, the authors of \cite{alzenad3D} developed an efficient algorithm for 3D UAV-BS placement to serve terrestrial users. Then, they extended their study in \cite{alzenadQoS} to the impact of the Quality-of-Service (QoS) irregularity on the optimal deployment of the UAV-BS. 
In \cite{kalantari2017user}, the authors proposed an efficient algorithm to maximize the sum-rate of terrestrial users by jointly optimizing the UAV-BSs wireless backhaul bandwidth and the wireless link association between terrestrial users and UAV-BSs. Also, authors in \cite{liu2019} investigated the joint placement and power allocation of a UAV-BS to maximize the coverage of terrestrial users. Finally, the authors of \cite{mozaffari2016efficient} studied the efficient deployment of multiple UAV-BSs for terrestrial coverage using the circle packing theorem.

Literature on cellular-connected UAVs, i.e., UAV-UEs, has been mainly focusing on reutilizing the terrestrial network as the exclusive communications provider to UAV-UEs.
For instance, authors of \cite{azari} derived the connectivity performance of a typical UAV-UE served by a network of terrestrial BSs, by leveraging tools from stochastic geometry. The same authors gave interesting insights and design recommendations in \cite{azari2} for UAV-UEs deployment under the assumption of exclusive cellular coverage provided by terrestrial networks.

Recently, we investigated in \cite{nesrineglobecom} the coverage probability of UAV-UEs in hybrid terrestrial/aerial networks, called vertical heterogeneous networks (VHetNets). We showed that in VHetNets, UAV-UEs are more likely to be served by UAV-BSs since terrestrial BSs do not provide enough power to UAV-UEs at high elevation angles. Moreover, the authors of  \cite{mozaffari2018beyond,mozaffari20183d} proposed a novel wireless network architecture of UAVs, where the optimal 3D cell association, i.e., between UAV-UEs and UAV-BSs, is developed in order to minimize the average latency at UAV-UEs. In their study, UAV-BSs were equipped with omni-directional antennas, and thus their 3D coverage was modeled as a truncated polyhedron. Yet, the 3D coverage of UAV-UEs using UAV-BSs equipped with directional antennas is still unexplored.

Consequently, we investigate in this paper the optimal 3D deployment of a directional-antenna equipped UAV-BS in an aerial system, where UAV-UEs are assumed to be exclusively associated with UAV-BSs. In particular, we
formulate the UAV-BS 3D placement problem aiming to maximize the number of covered UAV-UEs, given a spectrum sharing policy with terrestrial networks. Unlike terrestrial networks where all users are typically located at the ground altitude, UAV-UEs fly/hover in the 3D space at different locations and altitudes, with respect to flight regulations. As a consequence, the 3D coverage problem of UAV-UEs is more complex than in its terrestrial counterpart. 
Nevertheless, we propose a low complexity algorithm that determines the 3D optimal UAV-BS placement to cover the maximal number of UAV-UEs. 
Since spectrum sharing between aerial and terrestrial networks does affect the co-channel interference at terrestrial UEs, we propose a spectrum-aware UAV-BS 3D placement in the aerial network. The results obtained illustrate the coverage performance superiority of the UAV-BS under an orthogonal spectrum sharing (OSS) policy 
over its non-OSS (N-OSS) counterpart, where the UAV-BS placement is constrained by an interference-avoidance condition at the terrestrial network.
Also, they show distinct behaviors in OSS and N-OSS, given different parameters and configurations, e.g., antenna beamwidth, transmit power, aerial users density, aerial users clustering, and interference condition. Results obtained constitute essential guidelines to the co-existence of exclusive aerial networks with terrestrial networks.  

The rest of the paper is organized as follows. In Section II, the system model is presented. Section III formulates the 3D UAV-BS deployment problem under orthogonal and non-orthogonal spectrum sharing policies with terrestrial networks, and presents the associated solutions. Section IV illustrates the simulation results. Finally, Section V concludes the paper.
\vspace{-0.3cm}
\section{System Model}
\label{sec:sysmodel}
 \begin{figure}[t]
	\centering
	\includegraphics[width=0.9\linewidth]{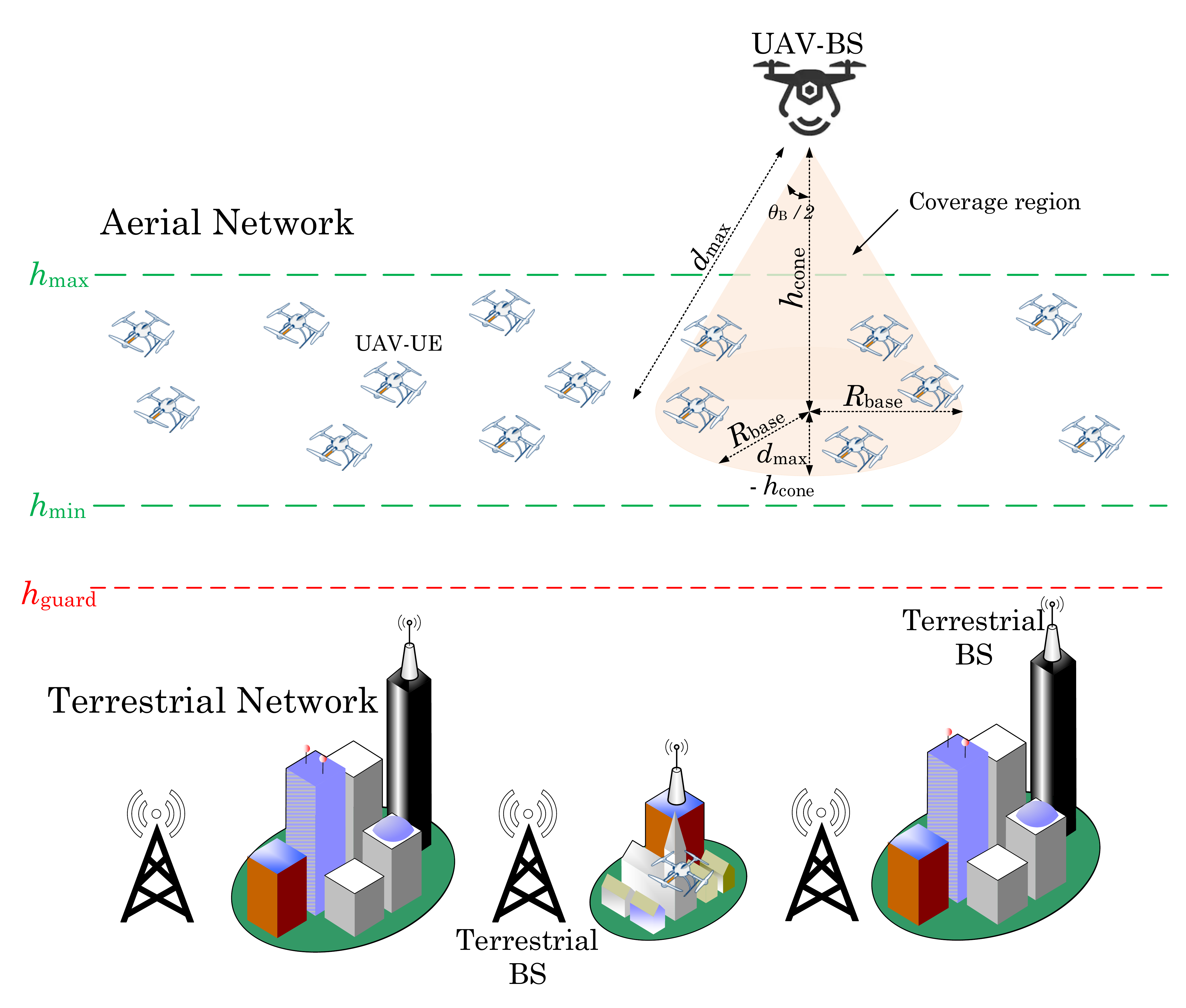}
	\caption{System model.}
	\label{fig:sys}
	  \vspace{-1.5em}
\end{figure}
We assume a 3D geographical area in the open sky, delimited by two altitude levels, namely $h_{\rm min}$ and $h_{\rm max}$, as shown in Fig. \ref{fig:sys}. In this area, a set $\mathcal{U}$ of UAV-UEs fly/hover at different altitudes. It is worth mentioning that according to the 3GPP report in \cite{3gpp777}, UAV-UEs cannot operate at altitudes above 300 meters. Moreover, the UAV-UEs are assumed to have the same power sensitivity $P_{\min}$, i.e., any UAV-UE succeeds in decoding a received signal when the latter's power is above $P_{\min}$. Let $\textbf{q}_i=[x_i,y_i,z_i]$ be the 3D location of the $i^{\rm th}$ UAV-UE belonging to the set $\mathcal{U}$, such that $h_{\rm min} \leq z_i \leq h_{\rm max}$.

We assume that a directional-antenna equipped UAV-BS will be deployed to serve the UAV-UEs. Its antenna gain (in dB) is given by \cite[eq (1)]{mozaffari2016efficient} 
\vspace{-0.2cm}
\begin{equation}
\label{eq:uavgain}
G =\left\{
\begin{array}{ll}
G_{\rm 3dB} ,& -\frac{\theta_{\rm B}}{2}\leqslant\psi\leqslant\frac{\theta_{\rm B}}{2},\\
g_{\rm s}, &\text{otherwise},
\end{array}
\right.
\end{equation}
 where $\frac{\theta_{\rm B}}{2}$ is the UAV-BS antenna's half beamwidth (in degrees), $\psi$ is the sector angle, $G_{\rm 3dB}\approx \frac{29000}{\theta_{\rm B}^2}$ is the main-lobe gain, and $g_{\rm s}$ is the side-lobe antenna gain. 
 
 In this work, we optimize the 3D location of the UAV-BS, designated by $\textbf{q}_{\rm{BS}}=[x_{\rm BS},y_{\rm BS},z_{\rm BS}]$, in order to serve the maximal number of UAV-UEs. Intuitively, the UAV-BS would be placed above the UAV-UEs such that its antenna is tilted-down towards them in order to provide cellular connectivity for their operations. Specifically, the altitude of the UAV-BS should be at least $z_{\rm BS} \geq h_{\rm max}$. Finally, we assume that the UAV-BS equivalent effective isotropic radiated power (EIRP), denoted $P_{\rm T}$, is written as 
 \vspace{-0.2cm}
 \begin{equation}
     \label{eq:ptmax}
     P_{\rm T}=P_t(\theta_{\rm B}) + G_{\rm 3dB},
     \end{equation}
where $P_t(\theta_{\rm B})$ is the transmit power associated to the beamwidth of the directional antenna. From (\ref{eq:uavgain}), we note that a larger antenna beamwidth $\theta_B$ results in a smaller 3 dB gain $G_{\rm{3dB}}$, thus $P_t(\theta_B)$ has to be increased in order to maintain a fixed $P_{\rm T}$.

When the UAV-BS communicates with the $i^{\rm th}$ UAV-UE, the received power can be expressed by 
\vspace{-0.2cm}
 \begin{equation}
     \label{eq:recPower}
     P_{i} =P_{\rm T} - \underbrace{10  n \log_{10} \left(\frac{4 \pi f_c d_i}{c}\right)}_{=L_{i}}  \geq P_{\min},
 \end{equation}
 where $n$ is the pathloss exponent, $f_c$ denotes the carrier frequency, $d_i$ is the distance between the $i^{\rm th}$ UAV-UE and the UAV-BS, and $L_{i}$ is the pathloss experienced by the $i^{\rm th}$ UAV-UE. The inequality in (\ref{eq:recPower}) ensures that the received power is above the sensitivity threshold. By taking the equality in (\ref{eq:recPower}), a pathloss maximal threshold can be defined as
 \vspace{-0.3cm}
\begin{equation}
    \label{eq:Pmin}
    L_{\rm th}=P_{\rm T}-P_{\min} \geq L_i.
     \vspace{-1em}
\end{equation} 

By substituting $L_i$ in (\ref{eq:Pmin}) by its expression in (\ref{eq:recPower}), and after some mathematical manipulations, we can define the 3D coverage region, denoted $\mathcal{C}_{\rm in}$, by a maximal distance between the UAV-BS and UAV-UEs, given by  
\vspace{-0.2cm}
 \begin{equation}
     \label{eq:dmax}
     d_{\max}= \frac{c}{4 \pi f_c} \; 10^{\frac{L_{\rm th}}{10 n}}.
 \end{equation}

Since aerial networks would co-exist with terrestrial networks, the interference component generated by the deployment of UAV-BSs and affecting the terrestrial users may be non-negligible, and hence requires particular adjustments of the aerial network. In this context, we distinguish two spectrum sharing policies between     
the aerial and terrestrial networks: 1) orthogonal spectrum sharing (OSS), where the UAV-BS communicates with its users at a different frequency band than the terrestrial BSs, and 2) non-OSS (N-OSS), where the aerial and terrestrial networks operate at the same frequency band. Under the OSS policy, terrestrial users won't experience interference from the UAV-BS, even if the latter operates at a relatively low altitude or high transmit power. However, under N-OSS, a regulatory altitude, called $h_{\rm guard}$, must be respected when deploying the UAV-BS, in order to prevent any undesired and additional interference at terrestrial users. $h_{\rm guard}$ can be seen as the highest altitude at which a terrestrial user can be located. Hence, the received interference power at any terrestrial user has to be lower than a certain threshold, called $\Delta$.\footnote{This power threshold can be seen as the peak tolerated interference by any terrestrial user, under the assumption that transmissions in the terrestrial network are continuous \cite{Jaafar2016}.} Specifically, we have
\vspace{-0.2cm}
\begin{equation}
    \label{eq:hguard}
    P_{\rm T} - 10 n \log \left(\frac{4 \pi f_c \left(z_{\rm BS}-h_{\rm guard}\right)}{c}\right) \leq \Delta,
\end{equation}
where $z_{\rm{BS}}-h_{\rm guard}$ is the smallest distance that can separate the UAV-BS from a terrestrial user, i.e., the latter is located directly below the UAV-BS at altitude $h_{\rm guard}$.
According to (\ref{eq:hguard}), the UAV-BS altitude has to respect the condition
\vspace{-0.1cm}
\begin{equation}
\label{eq:altitudeBS}
    z_{\rm{BS}} \geq \frac{c}{4 \pi f_c}\; 10^{\frac{P_{\rm T} - \Delta}{10 n}}+ h_{\rm{guard}}=\underbrace{d_{\max} \; 10^{\frac{P_{\min}-\Delta}{10 n}}}_{=d'_{\max}}+h_{\rm{guard}}.
\end{equation}

Based on (\ref{eq:altitudeBS}), the minimum altitude of the UAV-BS, when following a N-OSS policy, depends on both the receiver sensitivity of UAV-UEs and the tolerated interference level at terrestrial users.
\vspace{-0.2cm}
\subsection*{UAV-BS Coverage Region}
Based on (\ref{eq:Pmin})--(\ref{eq:dmax}), we infer that the 3D coverage region of the UAV-BS $\mathcal{C}_{\rm in}$ is composed of two volumes, as illustrated in Fig. \ref{fig:sys}:
\begin{itemize}
    \item A 3D cone with the UAV-BS placed at its tip. The cone's opening angle is $\theta_{\rm B}$ and its height, denoted $h_{\rm cone}$, is
    \vspace{-0.2cm}
    \begin{equation}
     \label{eq:hcone}
    h_{\rm cone}= d_{\max} \cos\left({\theta_{\rm B}}/{2}\right),
 \end{equation}
 whereas its base's radius, $R_{\rm base}$, can be written as
 \vspace{-0.2cm}
  \begin{equation}
     \label{eq:rbig}
    R_{\rm base}=  d_{\max} \sin\left({\theta_{\rm B}}/{2}\right).
 \end{equation}

\item A faced-down half ellipsoid  with radii $(R_{\rm base}, R_{\rm base}, d_{\max}-h_{\rm cone})$, and its center placed at coordinates $\textbf{q}_{\rm{elp}}=[x_{\rm BS}, y_{\rm BS}, z_{\rm{BS}}-h_{\rm cone}]$, where the ellipsoid equation is given by
\vspace{-0.2cm}
\begin{equation}
\frac{(x-x_{\rm BS})^2}{R_{\rm{base}}^2}+\frac{(y-y_{\rm BS})^2}{R_{\rm{base}}^2}+\frac{(z-z_{\rm{BS}}+h_{\rm{cone}})^2}{\left(d_{\max}-h_{\rm{cone}}\right)^2}=1. 
\end{equation}
\end{itemize}
 In the next section, we formulate the 3D placement problem, then we propose an algorithm that places the UAV-BS at the best location, which allows covering the maximal number of UAV-UEs. The 3D UAV-BS deployment depends on the adopted spectrum sharing policy, as it will be shown below.
 \section{Optimal UAV-BS 3D placement for maximum UAV-UEs coverage}
 Received interference at the terrestrial users from the UAV-BS depends on the latter's 3D placement and/or transmit power. Its deployment should be carefully planned, given the adopted spectrum sharing policy. In the subsequent sections, we investigate the 3D optimal placement under OSS and N-OSS policies.

 \subsection{UAV-BS 3D Placement Under OSS Policy}
 Under the OSS policy, The UAV-BS altitude $z_{\rm{BS}}$ is not restricted, as it doesn't result in any interference at the terrestrial users. 

\begin{remark}
The $i^{\rm th}$ UAV-UE falls into the  UAV-BS coverage region $\mathcal{C}_{\rm{in}}$ if it satisfies the two following conditions:
\vspace{-0.4cm}
\begin{subequations}
\begin{equation}
\label{condition1}
    c_1: \; \norm{\mathbf{q}_i-\mathbf{q}_{\rm{BS}}} \leq d_{\max}
\end{equation}
\begin{equation}
\label{condition2}
    c_2: \;   \cos{\left(\frac{\theta_B}{2}\right)} \leq \frac{|z_i- z_{\rm{BS}}|}{\norm{\mathbf{q}_i-\mathbf{q}_{\rm{BS}}}},
\end{equation}
\end{subequations}
where $\norm{\cdot}$ is the Euclidean norm and $|\cdot|$  is the absolute value operations.
\end{remark}
According to (\ref{condition1}), the $i^{\rm th}$ UAV-UE is within the region $\mathcal{C}_{\rm{in}}$ when it is distant from the UAV-BS by at most $d_{\max}$. 
Meanwhile, (\ref{condition2})
ensures that the elevation angle between the $i^{\rm th}$ UAV-UE and UAV-BS is below the UAV-BS antenna half-beamwidth  $\frac{\theta_{\rm B}}{2}$.

Let $c_i$ be the binary variable that indicates whether the $i^{\rm th}$ UAV-UE is in $\mathcal{C}_{\rm in}$ or not, i.e., $c_i=1$ when it is in the coverage region of the UAV-BS, and $c_i=0$ otherwise. 
Consequently, the UAV-BS coverage maximization problem can be formulated as a mixed integer non-linear problem (MINLP) as follows:
\vspace{-0.2cm}
\begin{subequations}
	\begin{align}
	\max_{\mathbf{q}_{\rm{BS}}, \mathbf{c}}   & \quad      \sum_{i \in \mathcal{U}} c_i \tag{P1} 
	\\
	\label{c2_21} \text{s.t.}\quad & c_i\times \norm{\mathbf{q}_i-\mathbf{q}_{\rm{BS}}} \leq d_{\max}, \forall i \in \mathcal{U},  \nonumber \tag{P1.a}\\
	\label{c2_2} &c_i\times{\norm{\mathbf{q}_i-\mathbf{q}_{\rm{BS}}}} \leq \frac{|z_i- z_{\rm{BS}}|}{\cos{\left(\frac{\theta_{\rm B}}{2}\right)}}, \forall i \in \mathcal{U}, \tag{P1.b}\\
	\label{c2_3} & c_i \in \{0,1\}, \forall i \in \mathcal{U}, \tag{P1.c}
	\end{align}
\end{subequations}
where $\textbf{c}=[c_1,\ldots,c_{U}]$, and $U$ is the total number of UAV-UEs. In order to guarantee the feasibility of the constraints (\ref{c2_21})--(\ref{c2_2}) when $c_i=0$, we rewrite them as follows \cite{alzenad3D}:
\vspace{-0.2cm}
\begin{subequations}
\begin{equation}
\label{condition1_1}
    c_1: \; \norm{\mathbf{q}_i-\mathbf{q}_{\rm{BS}}} \leq d_{\max} + N (1-c_i),\; \forall i \in \mathcal{U}
\end{equation}
\begin{equation}
\label{condition2_1}
    c_2: \;   {\norm{\mathbf{q}_i-\mathbf{q}_{\rm{BS}}}} \leq \frac{|z_i- z_{\rm{BS}}|}{\cos{\left(\frac{\theta_{\rm B}}{2}\right)}}- N (1- c_i), \; \forall i \in \mathcal{U},
\end{equation}
\end{subequations}
where $N$ is a very large number. 
The complexity of solving (P1) arises from the dependency between $\mathbf{q}_{\rm{BS}}$ and $z_{\rm BS}$ in constraint (\ref{condition2_1}). To tackle this issue, we propose to decouple the 2D UAV-BS location and its altitude's optimization. Indeed, for a specific UAV-BS altitude $z_{\rm BS}^1$, (P1) reduces to a 2D UAV-BS location optimization problem. Then, exhaustive search over the altitude allows to find the optimal solution to (P1).
The 2D optimization problem can be written as
\vspace{-0.2cm}
\begin{subequations}
	\begin{align}
	\max_{{x}_{\rm{BS}}, {y}_{\rm{BS}},  \mathbf{c}}   & \quad      \sum_{i \in \mathcal{U}} c_i \tag{P2} 
	\label{cc1}\\
	\text{s.t.}\quad & 
	\label{cc11}\norm{\mathbf{q}_i-\mathbf{q}_{\rm{BS}}^1} \leq d_{\max}+N (1- c_i),\forall i \in \mathcal{U}
	\nonumber \tag{P2.a}\\
	\label{cc2} & {\norm{\mathbf{q}_i-\mathbf{q}_{\rm{BS}}^1}}  \leq \frac{|z_i- z_{\rm{BS}}^1|}{\cos{\left(\frac{\theta_{\rm B}}{2}\right)}}-N (1-c_i), \forall i \in \mathcal{U} \tag{P2.b}\\
	\label{cc3} & c_i \in \{0,1\}, \forall i \in \mathcal{U}, \tag{P2.c}
	\end{align}
\end{subequations}
where $\textbf{q}_{\rm{BS}}^1=[x_{\rm{BS}}, y_{\rm{BS}}, z_{\rm{BS}}^1]$. 
Since (\ref{cc11})-(\ref{cc2}) constraints are convex, (P2) can be solved using MOSEK parser in the CVX package of Matlab \cite{CVX}. 
\begin{algorithm}[t]
\caption{3D UAV-BS placement under OSS policy}
\label{Algo1}
\begin{algorithmic}[1]
\Require{$P_{\rm T},\theta_{\rm B},P_{\rm min}, f_c, h_{\min},h_{\max}$}
\Ensure{($\textbf{q}_{\rm BS}^{\rm opt}$, $\textbf{c}^{\rm opt}$)}
 \State Compute $d_{\max}$ using (\ref{eq:dmax})
\For{$z_{\rm BS}^1 \in [h_{\max},h_{\max}+d_{\max}]$}
        \State \text{Solve (P2)}, get ($\bar{\textbf{q}}_{\rm BS}^1, \bar{\textbf{c}}^1$), and store it in a vector $\textbf{v}$
      \EndFor
      \State Select the best element in $\textbf{v}$ that maximizes $\sum_{i \in \mathcal{U}} c_i$
      \State Return ($\textbf{q}_{\rm BS}^{\rm opt}$, $\textbf{c}^{\rm opt}$) as the best element in $\textbf{v}$.
\end{algorithmic}
\end{algorithm}
Assuming that (${\bar{\textbf{q}}}_{\rm{BS}}^1$, $\bar{\textbf{c}}^1$) is the optimal solution of (P2) for a given $z_{\rm{BS}}^1$, then (P1) can be solved iteratively using Algorithm \ref{Algo1}, detailed above. In line 2 of Algorithm \ref{Algo1}, $z_{\rm BS}^1$ is in the range $[h_{\max}, h_{\max}+d_{\max}]$. Without loss of generality, $d_{\max}\geq h_{\max}-h_{\min}$, thus $z_{\rm BS}^1=h_{\max}$ ensures that the vertical coverage of the UAV-BS is deep enough to cover even UAV-UEs flying/hovering at the lowest authorized altitude $h_{\min}$. In contrast, when $z_{\rm BS}^1=h_{\max}+d_{\max}$, the coverage area of the UAV-BS is tangent to the horizontal plan at altitude $h_{\max}$, i.e., only a few users flying/hovering at $h_{\max}$ can be covered.   
\vspace{-0.1cm}
 \subsection{UAV-BS 3D Placement Under N-OSS Policy}

Under the N-OSS policy, the UAV-BS coverage region is restricted in order to respect the tolerated interference level at terrestrial users.
In this case, the UAV-BS' total transmit power, $P_{\rm T}$, should be carefully adjusted to ensure that the UAV-BS hovers at an altitude of at least $h_{\max}$\footnote{Being at least at $h_{\max}$ with a sufficient coverage depth, $d_{\max}$, ensures a maximized vertical coverage.} as described in Fig. \ref{fig:sys}. Thus, the UAV-BS total transmit power should be no smaller than $P_{\rm T}^{\rm low}$ expressed by 
\vspace{-0.2cm}
\begin{equation}
    \label{eq:minPtmax}
    P_{\rm T}^{\rm low}=10 n \log \left(\frac{4 \pi f_c \left(h_{\max}-h_{\rm guard}\right)}{c}\right) +\Delta.
\end{equation}
Since the UAV-BS altitude cannot exceed $d_{\max}+h_{\max}$\footnote{Above this altitude, the coverage performance is null since no UAV-UE flies/hovers at altitudes higher than $h_{\max}$.}, the UAV-BS total transmit power cannot exceed $P_{\rm T}^{\rm high}$, which is given by 
\vspace{-0.3cm}
\begin{equation}
    \label{eq:ptmaxth}
    P_{\rm T}^{\rm high}=10 n \log \left(\frac{4 \pi f_c \left(h_{\max}-h_{\rm guard}\right)}{c \times \Omega}\right),
\end{equation}
where $\Omega=10^{-\frac{\Delta}{10n}}-10^{-\frac{P_{\min}}{10n}}$.
\begin{proof}
The above result follows from substituting (\ref{eq:Pmin})--(\ref{eq:dmax}) into (\ref{eq:altitudeBS}), where equality is taken for $z_{\rm BS}=d_{\max}+h_{\max}$.
\end{proof}
Finally, using (\ref{eq:minPtmax})--(\ref{eq:ptmaxth}), we have $P_{\rm T}^{\rm low} \leq P_{\rm T} \leq P_{\rm T}^{\rm high}$.

The UAV-BS 3D placement problem in N-OSS can be formulated as in (P1) with the additional optimization parameter $P_{\rm T}$ and constraints (\ref{eq:altitudeBS}), (\ref{eq:minPtmax}), and (\ref{eq:ptmaxth}). Unlike OSS policy, the UAV-BS EIRP should be optimized to guarantee two conditions: 1) the UAV-BS coverage region is maximized, and 2) imposed spectrum-sharing interference condition is respected.
Subsequently, a similar approach as in Algorithm \ref{Algo1} can be followed to solve it. The proposed approach is presented in Algorithm \ref{Algo2}, where $P_{\rm T}^1 \in [P_{\rm T}^{\rm low}, P_{\rm T}^{\rm high}]$ and $z_{\rm BS}^1 \in [h_{\max}, d_{\max}+ h_{\max}]$. 

\begin{algorithm}[t]
\caption{3D UAV-BS placement under N-OSS policy}
\label{Algo2}
\begin{algorithmic}[1]
\Require{$\theta_{\rm B}, P_{\rm min}, f_c, h_{\min}, h_{\max}, h_{\rm guard}, \Delta$} 
\Ensure{($\textbf{q}_{\rm BS}^{\rm opt}$, $\textbf{c}^{\rm opt}$, ${P}_{\rm T}^{\rm opt}$)}
 \State Compute $P_{\rm T}^{\min}$ and $P_{\rm T}^{\rm th}$ using (\ref{eq:minPtmax}) and (\ref{eq:ptmaxth}) resp.
\For{$P_{\rm T}^1 \in [P_{\rm T}^{\rm low},P_{\rm T}^{\rm high}]$}
\State Compute $d_{\max}$ and $d'_{\max}$ using (\ref{eq:dmax}) and (\ref{eq:altitudeBS}) resp.
\For{$z_{\rm BS}^1 \in [h_{\max},h_{\max}+d_{\max}]$}
        \State \text{Solve (P2)}, get ($\bar{\textbf{q}}_{\rm BS}^1, \bar{\textbf{c}}^1$), and store it in vector $\textbf{v}$
      \EndFor
       \State Select the best element in $\textbf{v}$ that maximizes $\sum_{i \in \mathcal{U}} c_i$ and store it with $P_{\rm T}^1$ in a vector $\textbf{u}$
      \EndFor
      \State Select the best element in $\textbf{u}$ that maximizes $\sum_{i \in \mathcal{U}} c_i$
      \State Return ($\textbf{q}_{\rm BS}^{\rm opt}$, $\textbf{c}^{\rm opt}$, ${P}_{\rm T}^{\rm opt}$) as the best element in $\textbf{u}$.
\end{algorithmic}
\end{algorithm}

\section{Simulation Results}
In our simulations, we consider a 3 km × 3 km area. Unless stated otherwise, we assume that $P_{\rm T}=30$ dBm, $f_c=2$ GHz, $P_{\min}=-70$ dBm, $\theta_{\rm B}=60^{\circ}$, $h_{\min}=100$ meters, $h_{\max}=300$ meters, $h_{\rm guard}=50$ meters, and $\Delta=-73$ dBm. Also, simulations are averaged for 500 UAV-UEs scenarios. For the UAV-UEs distribution per volume unit (${\rm km}^3$), we assume two point processes:
\begin{itemize}
    \item Homogeneous Poisson point process (HPPP) for UAV-UEs uniform distribution in the 3D area delimited vertically by $h_{\min}$ and $h_{\max}$, with density $\lambda$ UAV-UEs/km$^3$.
    \item Mat\'{e}rn cluster point process (MCPP) requiring two overlaying point processes: 1) a HPPP with density $\lambda_P$/${\rm km}^3$ as the \textit{parent} cluster heads, and 2) a \textit{daughter} HPPP with density $\lambda_D$ of UAV-UEs belonging to a sphere with radius $r_{D}$ centered at each \textit{parent} cluster head. MCPP mimics the heterogeneity of the UAV-UEs distribution, where a higher $\lambda_D$ means more clustered UAV-UEs.
\end{itemize}

\begin{figure*}[!t]
\centering
	\subfigure[Avg. max. number of covered UAV-UEs vs. UAV-BS altitude.]{
		\includegraphics[scale=0.29]{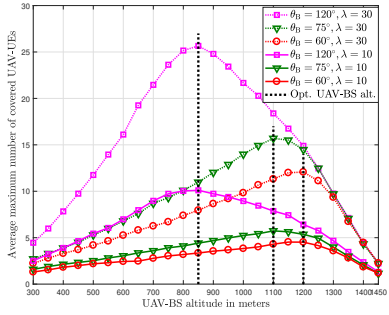}
	\label{fig:height}
	}
	\hfill
	\subfigure[Avg. max. number of covered UAV-UEs vs. UAV-BS antenna beamwidth.]{
	\includegraphics[scale=0.29]{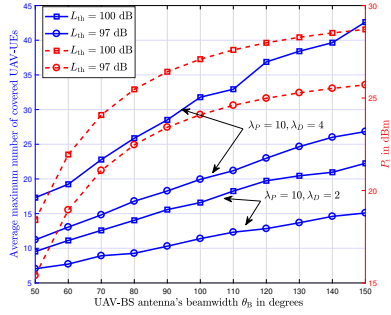}
		\label{fig:nbBeam}
	}
	\hfill
	\subfigure[Avg. max. number of covered UAV-UEs vs. distribution of UAV-UEs.]{
	\includegraphics[scale=0.26]{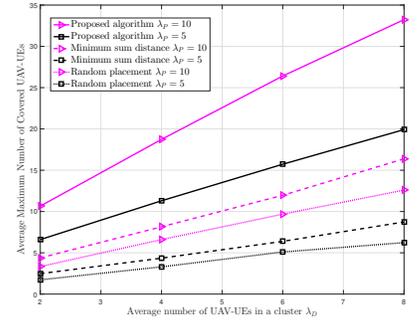}
		\label{fig:nbMatern}
	}
	\caption[]{UAV-BS coverage performance under OSS policy.}
	\label{Fig:OSS}
\end{figure*}

\subsection{UAV-BS Coverage Performance Under OSS Policy}

Fig. \ref{Fig:OSS} presents the optimal coverage performance of the UAV-BS under OSS policy. 

Fig. \ref{fig:height} shows the average maximum number of covered UAV-UEs (a.k.a. coverage performance) as a function of UAV-BS altitude $z_{\rm BS}^1$, for different $\theta_B$ and HPPP $\lambda$ values. Under OSS, an optimal altitude can be reached, which does not depend on the distribution of the UAV-UEs $\lambda$. However, as the UAV-BS antenna's beamwidth $\theta_B$ increases, the UAV-BS hovers at a lower altitude to cover more UAV-UEs.

In Fig. \ref{fig:nbBeam}, the impact of the UAV-BS antenna's beamwidth $\theta_{B}$, for different $L_{\rm th}$ is investigated, given an MCPP UAV-UEs distribution. As $\theta_{B}$ or $L_{\rm th}$ increases, the coverage performance improves. The same occurs when $\lambda_D$ is high, i.e., UAV-UEs clusters are denser. 
Moreover, the transmit power $P_t(\theta_B)$ of the UAV-BS is evaluated using (\ref{eq:ptmax}). When $\theta_B$ is higher, $G_{\rm 3dB}$ decreases, thus increasing $P_t(\theta_B)$, i.e., covering more UAV-UEs using a wide-beamwidth antenna comes at the expense of a higher transmit power.

Fig. \ref{fig:nbMatern} compares the coverage performance of our proposed algorithm to that of two benchmarks, namely ``minimum sum distance" proposed in \cite{liu2019} where the UAV-BS placement is determined for minimum pathloss to all UAV-UEs, 
and ``random placement" where the UAV-BS location is randomly selected. The coverage performance is evaluated as a function of the MCPP density $\lambda_D$, and for different MCPP $\lambda_P$ values.
We found that UAV-UEs clustering, i.e., higher $\lambda_D$ and $\lambda_P$, improves the coverage performance for all UAV-BS placement approaches. Nevertheless, our algorithm significantly outperforms both benchmarks.
\vspace{-0.1in}
\subsection{UAV-BS Coverage Performance Under N-OSS Policy}
\begin{figure*}[!t]
\centering
	\subfigure[Avg. max. number of covered UAV-UEs vs. UAV-BS altitude.]{
		\includegraphics[scale=0.31]{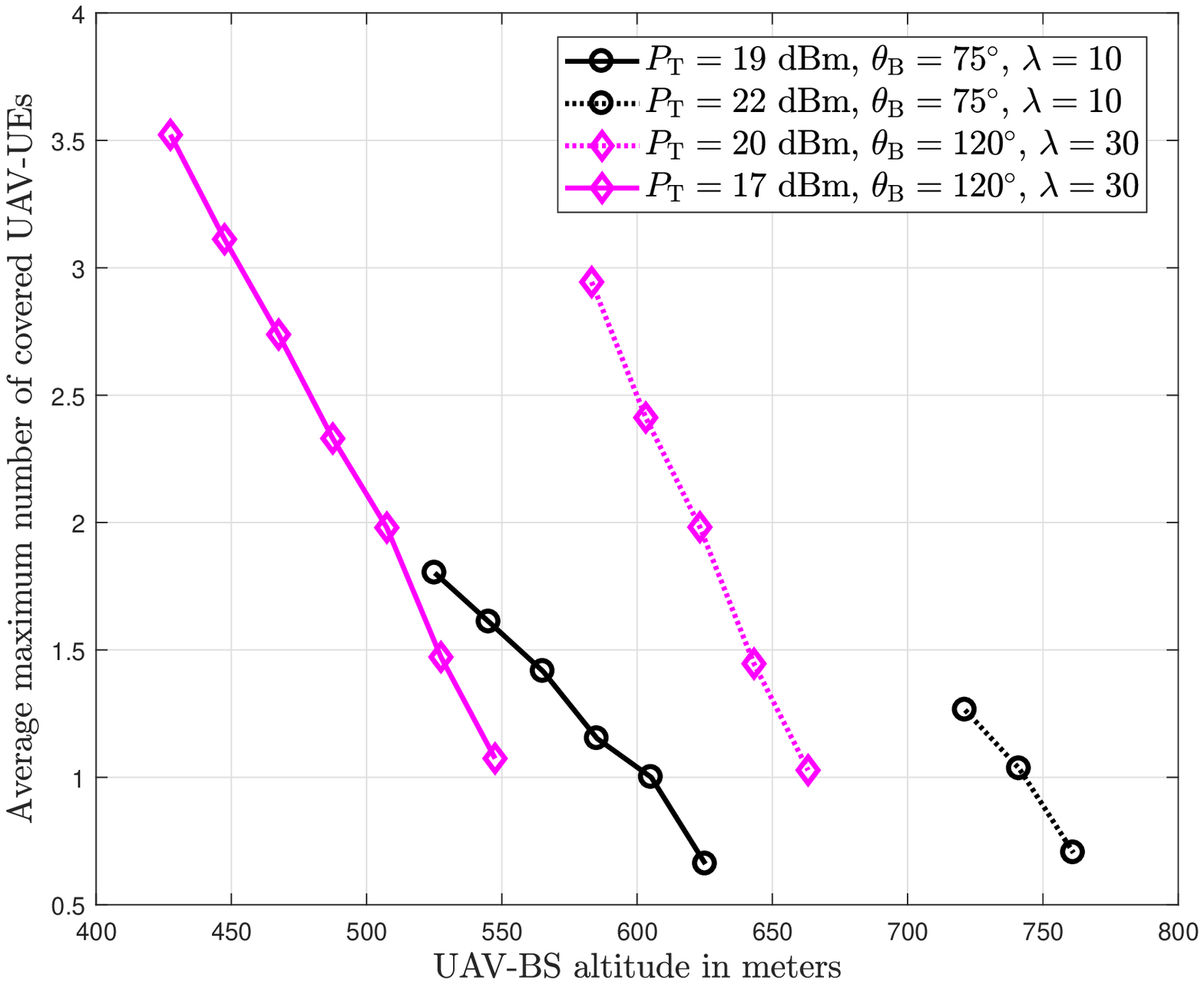}
	\label{fig:heightNOSS}
	}
	\hfill
	\subfigure[Avg. max. number of covered UAV-UEs vs. UAV-BS total transmit power.]{
	\includegraphics[scale=0.31]{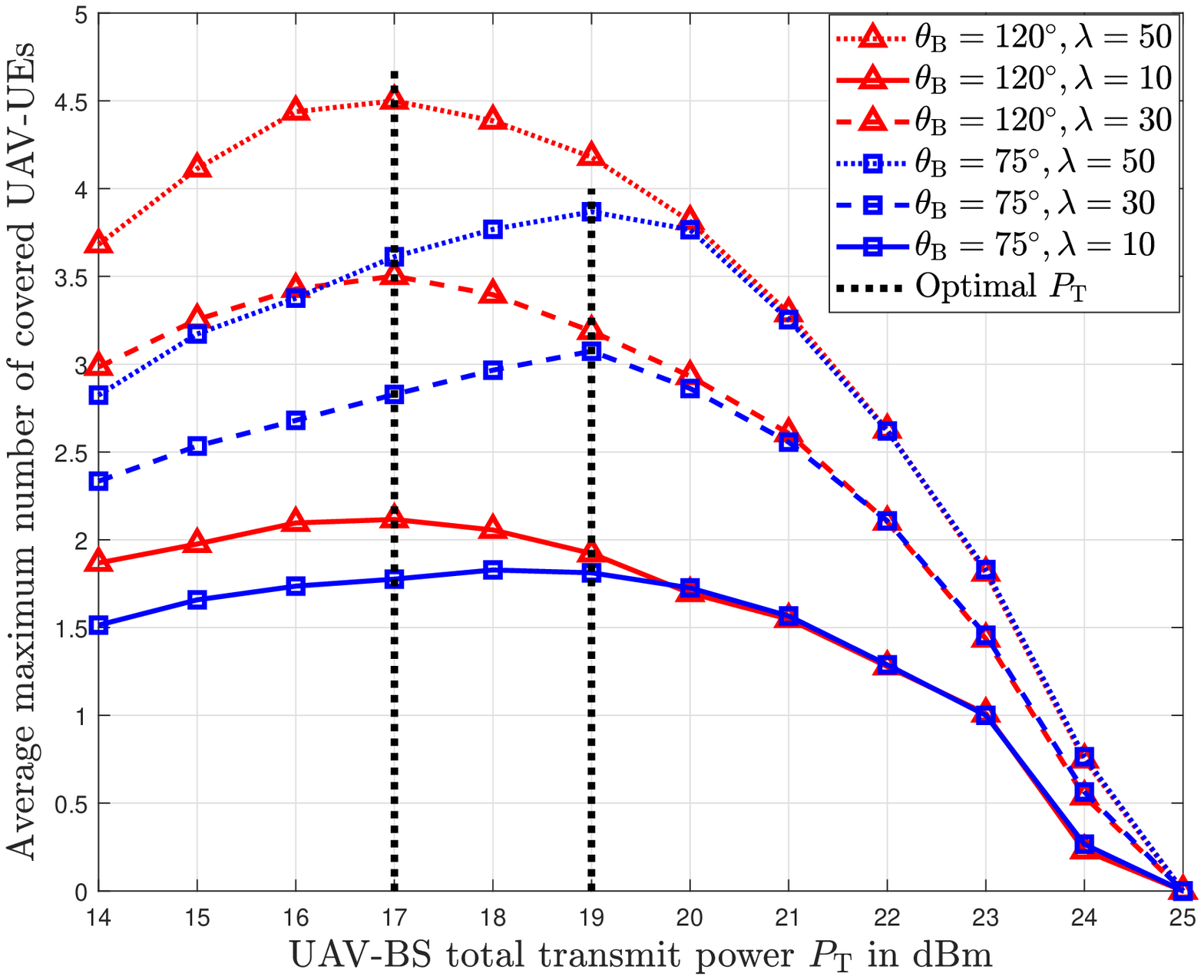}
		\label{fig:PtNOSS}
	}
	\hfill
	\subfigure[Avg. max. number of covered UAV-UEs vs. the N-OSS regulatory altitude.]{
	\includegraphics[scale=0.32]{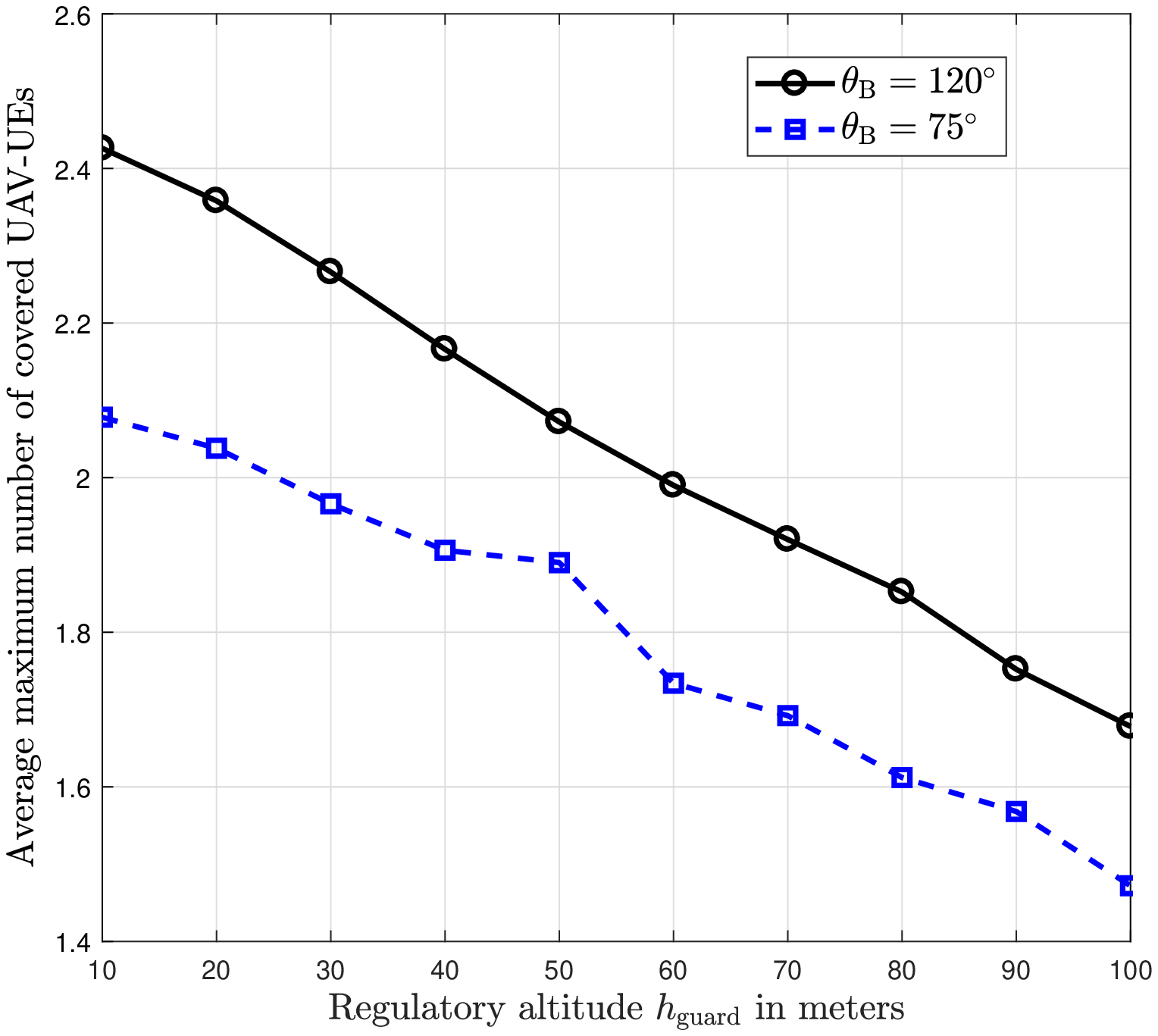}
		\label{fig:hguardNOSS}
	}
	\caption[]{UAV-BS coverage performance under N-OSS policy.}
	\label{Fig:NOSS}
\end{figure*}

Fig. \ref{Fig:NOSS} presents the optimal coverage performance of the UAV-BS under N-OSS policy. 

In Fig. \ref{fig:heightNOSS}, we show the coverage performance as a function of the UAV-BS altitude, given a HPPP UAV-UEs distribution.
For any ($P_{\rm T}$, $\theta_B$, $\lambda$) setup, the optimal altitude is always the smallest allowed one, which is given by (\ref{eq:altitudeBS}).
This is expected since increasing the altitude can only reduce the overlapping area between the coverage region and the operating UAV-UEs' air corridor, thus reducing the number of covered UAV-UEs.
This interesting outcome allows to reduce the steps taken in Algorithm \ref{Algo2} by omitting step 4, and calculating the best altitude for a given $P_{\rm T}$ using (\ref{eq:altitudeBS}) instead.

Fig. \ref{fig:PtNOSS} studies the impact of $P_{\rm T}$ on the coverage performance, for different $\theta_B$ and $\lambda$. For any ($\theta_B$, $\lambda$) setup, an optimal $P_{\rm T}$ value can be obtained, which maximizes the coverage performance.  
Also, as $\theta_B$ reduces, i.e., narrower antenna beamwidth, more transmit power is required to cover the maximal number of UAV-UEs. Indeed, the loss in horizontal coverage (wider $\theta_B$) is compensated by a deeper vertical coverage (higher $P_{\rm T}$), with respect to the interference condition $\Delta$ (i.e., moving to a higher altitude as shown in Fig. \ref{fig:heightNOSS}). 
Therefore, a low-altitude UAV-BS uses less transmit power for connectivity to UAV-UEs, while keeping its interference at the terrestrial users lower than $\Delta$.

Finally, the impact of the regulatory altitude, $h_{\rm guard}$, on the coverage performance is illustrated in Fig. \ref{fig:hguardNOSS}. A more relaxed regulatory altitude for terrestrial users, i.e., smaller $h_{\rm guard}$, increases the average maximum number of covered UAV-UEs. Indeed, the UAV-BS would move with more flexibility in order to increase the overlapping area between its coverage region and the UAV-UEs' air corridor. Furthermore, wider UAV-BS antenna beamwidth always provides a better coverage performance independently from the regulatory altitude.
Eventually, under the N-OSS policy, the regulatory altitude $h_{\rm guard}$ and the peak tolerated interference at terrestrial users $\Delta$ should be carefully considered as they significantly impact the 3D UAV-BS placement and coverage performance.

\vspace{-0.05in}
\section{Conclusion}
  In this paper, we studied the 3D placement problem of a directional-antenna equipped UAV-BS, aiming to maximize the number of flying/hovering UAV-UEs under its coverage region. The problem is formulated under two spectrum sharing policies, OSS and N-OSS, where frequency resources are either shared orthogonally or non-orthogonally between the aerial and terrestrial networks. 
Then, we solved the 3D UAV-BS placement using an iterative approach, where the horizontal 2D location of the UAV-BS is optimized for each explored altitude, and finally the altitude demonstrating the best coverage performance is selected.
Simulation results have shown that under OSS policy, the UAV-BS hovers at low altitudes when its antenna's beamwidth is large. Also, we found that differences in UAV-UEs location distributions affect the performance of the UAV-BS. Indeed, the coverage performance improves when more UAV-UEs are clustered together. On the other hand, under N-OSS policy, the optimal UAV-BS altitude is always the lowest authorized one. Moreover, UAV-BS transmit power has to be optimized for maximum 3D coverage, while satisfying the interference condition on the terrestrial network. Finally, N-OSS results showed that a narrower UAV-BS antenna beamwidth is compensated by a higher transmit power and altitude.

\vspace{-0.3cm}
\bibliographystyle{IEEEtran}  
\bibliography{references}  

\begin{thebibliography}{10}
\providecommand{\url}[1]{#1}
\csname url@rmstyle\endcsname
\providecommand{\newblock}{\relax}
\providecommand{\bibinfo}[2]{#2}
\providecommand\BIBentrySTDinterwordspacing{\spaceskip=0pt\relax}
\providecommand\BIBentryALTinterwordstretchfactor{4}
\providecommand\BIBentryALTinterwordspacing{\spaceskip=\fontdimen2\font plus
\BIBentryALTinterwordstretchfactor\fontdimen3\font minus
  \fontdimen4\font\relax}
\providecommand\BIBforeignlanguage[2]{{%
\expandafter\ifx\csname l@#1\endcsname\relax
\typeout{** WARNING: IEEEtran.bst: No hyphenation pattern has been}%
\typeout{** loaded for the language `#1'. Using the pattern for}%
\typeout{** the default language instead.}%
\else
\language=\csname l@#1\endcsname
\fi
#2}}

\bibitem{Faawebsite}
\BIBentryALTinterwordspacing
{Federal Aviation Administration (FAA)}. [Online]. Available:
  \url{https://www.faa.gov/uas/resources/by_the_numbers/}
\BIBentrySTDinterwordspacing

\bibitem{irem}
I.~{Bor-Yaliniz} and H.~{Yanikomeroglu}, ``The new frontier in {RAN}
  heterogeneity: Multi-tier drone-cells,'' \emph{{IEEE Commun. Mag.}}, vol.~54,
  no.~11, pp. 48--55, Nov. 2016.

\bibitem{mozaffari2018beyond}
M.~Mozaffari, A.~T.~Z. Kasgari, W.~Saad, M.~Bennis, and M.~Debbah, ``Beyond
  {5G} with {UAV}s: Foundations of a {3D} wireless cellular network,''
  \emph{{IEEE Trans.\ Wireless Commun.}}, vol.~18, no.~1, pp. 357--372, Jan.
  2019.

\bibitem{3gppTS22125}
3GPP, ``Unmanned aerial system {(UAS)} support in {3GPP} (release 17), {TS}
  22.125,'' Dec. 2019.

\bibitem{3gpp777}
------, ``Study on enhanced {LTE} support for aerial vehicles (release 15),
  {TR} 36.777,'' Jun. 2018.

\bibitem{nesrineglobecom}
N.~{Cherif}, M.~{Alzenad}, H.~{Yanikomeroglu}, and A.~{Yongacoglu}, ``Downlink
  coverage analysis of an aerial user in vertical heterogeneous networks,'' in
  \emph{{Proc. IEEE Glob. Commun. Conf.~(Globecom)}}, Waikoloa, HI, USA, Dec.
  2019, pp. 1--6.

\bibitem{alzenad3D}
M.~Alzenad, A.~El-Keyi, F.~Lagum, and H.~Yanikomeroglu, ``{3-D} placement of an
  unmanned aerial vehicle base station {(UAV-BS)} for energy-efficient maximal
  coverage,'' \emph{{IEEE Wireless Commun.\ Lett.}}, vol.~6, no.~4, pp.
  434--437, Aug. 2017.

\bibitem{alzenadQoS}
M.~Alzenad, A.~El-Keyi, and H.~Yanikomeroglu, ``{3-D} placement of an unmanned
  aerial vehicle base station for maximum coverage of users with different
  {QoS} requirements,'' \emph{{IEEE Wireless Commun.\ Lett.}}, vol.~7, no.~1,
  pp. 38--41, Feb. 2018.

\bibitem{kalantari2017user}
E.~Kalantari, I.~Bor-Yaliniz, A.~Yongacoglu, and H.~Yanikomeroglu, ``User
  association and bandwidth allocation for terrestrial and aerial base stations
  with backhaul considerations,'' in \emph{{Proc. IEEE Int. Symp. on Pers.,
  Indoor and Mobile Radio Commun.~(PIMRC)}}, Montreal, QC, Canada, Oct. 2017,
  pp. 1--6.

\bibitem{liu2019}
X.~{Liu}, J.~{Wang}, N.~{Zhao}, Y.~{Chen}, S.~{Zhang}, Z.~{Ding}, and F.~R.
  {Yu}, ``Placement and power allocation for {NOMA-UAV} networks,'' \emph{{IEEE
  Wireless Commun.\ Lett.}}, vol.~8, no.~3, pp. 965--968, Jun. 2019.

\bibitem{mozaffari2016efficient}
M.~Mozaffari, W.~Saad, M.~Bennis, and M.~Debbah, ``Efficient deployment of
  multiple unmanned aerial vehicles for optimal wireless coverage,''
  \emph{{IEEE Commun.\ Lett.}}, vol.~20, no.~8, pp. 1647--1650, Aug. 2016.

\bibitem{azari}
M.~M. Azari, F.~Rosas, A.~Chiumento, and S.~Pollin, ``Coexistence of
  terrestrial and aerial users in cellular networks,'' in \emph{{Proc. IEEE
  Glob. Commun. Conf.~(Globecom) Workshops}}, Singapore, Dec. 2017, pp. 1--6.

\bibitem{azari2}
M.~M. {Azari}, F.~{Rosas}, and S.~{Pollin}, ``Cellular connectivity for {UAVs}:
  Network modeling, performance analysis, and design guidelines,'' \emph{IEEE
  Trans. Wireless Commun.}, vol.~18, no.~7, pp. 3366--3381, Jul. 2019.

\bibitem{mozaffari20183d}
M.~Mozaffari, A.~T.~Z. Kasgari, W.~Saad, M.~Bennis, and M.~Debbah, ``{3D}
  cellular network architecture with drones for beyond {5G},'' in \emph{{Proc.
  IEEE Glob. Commun. Conf.~(Globecom)}}, Abu Dhabi, UAE, Dec. 2018, pp. 1--6.

\bibitem{Jaafar2016}
W.~{Jaafar}, T.~{Ohtsuki}, W.~{Ajib}, and D.~{Haccoun}, ``{Impact of the CSI on
  the performance of cognitive relay networks with partial relay selection},''
  \emph{IEEE Trans. Veh. Tech.}, vol.~65, no.~2, pp. 673--684, Feb. 2016.

\bibitem{CVX}
M.~Grant and S.~P. Boyd, ``{CVX: MATLAB software for disciplined convex
  programming},'' \emph{http://cvxr.com/cvx/}, Jan. 2014.

\end{thebibliography}

\end{document}